\documentclass[draftcls,onecolumn,12pt]{IEEEtran}
\usepackage{bbding}
\usepackage{multirow}
\usepackage{amsfonts}
\usepackage{booktabs}
\usepackage{colortbl}
\usepackage{amsfonts}
\usepackage{mathrsfs}
\usepackage{cite}
\usepackage{graphicx}
\usepackage{amsmath}
\usepackage{amssymb}
\usepackage{stfloats}
\usepackage{array}
\usepackage{times}

\newcolumntype{I}{!{\vrule width 1pt}}

\newlength\savewidth
\newcommand{\PreserveBackslash}[1]{\let\temp=\\#1\let\\=\temp}
\newcolumntype{C}[1]{>{\PreserveBackslash\centering}p{#1}}
\newcolumntype{R}[1]{>{\PreserveBackslash\raggedleft}p{#1}}
\newcolumntype{L}[1]{>{\PreserveBackslash\raggedright}p{#1}}
\newcommand\shline{\noalign{\global\savewidth\arrayrulewidth
                            \global\arrayrulewidth 1pt}%
                   \hline
                   \noalign{\global\arrayrulewidth\savewidth}}

\begin{document}

\title{Simultaneous Bidirectional Link Selection in Full Duplex MIMO Systems}

\author{\IEEEauthorblockN{Mingxin
Zhou\IEEEauthorrefmark{1}, Lingyang Song\IEEEauthorrefmark{1}, Yonghui Li\IEEEauthorrefmark{2}, and Xuelong Li\IEEEauthorrefmark{3}}
\IEEEauthorblockA{\\\IEEEauthorrefmark{1}
School of Electronics Engineering and Computer Science, Peking University, Beijing, China. $100871$\\
\IEEEauthorrefmark{2}Centre of Excellence in Telecommunications, School of Electrical and Information Engineering, The University of Sydney, Australia\\\IEEEauthorrefmark{3}State Key Laboratory of Transient Optics and Photonics, Xi'an Institute of Optics and Precision Mechanics, Chinese Academy of Sciences, Xi'an 710119, Shaanxi, P. R. China\\
}}
\maketitle
\begin{abstract}
In this paper, we consider a point to point full duplex (FD) MIMO
communication system. We assume that each node is equipped with
an arbitrary number of antennas which can be used for transmission or reception.
With FD radios, bidirectional information exchange between two
nodes can be achieved at the same time. In this paper we design
bidirectional link selection schemes by selecting a pair of transmit and receive antenna at both ends for communications in each direction to maximize the weighted sum rate or minimize the weighted sum symbol error rate (SER). The optimal selection schemes require exhaustive search, so they are highly complex. To tackle this problem, we propose a Serial-Max selection algorithm, which approaches the exhaustive search methods with much lower complexity. In the Serial-Max method, the antenna pairs with
maximum ``obtainable SINR'' at both ends are selected in a two-step serial way. The performance of the
proposed Serial-Max method is analyzed, and the closed-form
expressions of the average weighted sum rate and the weighted sum SER are derived. The
analysis is validated by simulations. Both analytical and simulation
results show that as the number of antennas increases, the Serial-Max method approaches
the performance of the exhaustive-search schemes in terms of sum rate and sum SER.

\end{abstract}


\IEEEpeerreviewmaketitle
\section{Introduction}

Current wireless communications systems typically exploit half
duplex (HD) transmission. This is because for many years the full duplex (FD)
transmission has been considered impractical. The signal leakage
from the local output to input in the FD radio, referred to as the self interference,
may overwhelm the receiver, thus making it impossible to extract
the desired signals. Very recently, there has been a significant
process in self-interference suppression in FD radios. The passive
suppression methods~\cite{Ref:J.C:ASC, Ref:Empower, Ref:rethink,
Ref:Bid_MIT,Ref:null,Ref:beam,Ref:exp,Ref:spa_null} design the
antennas using a combination of path loss, cross polarization and
antenna directionality, while the active
approaches~\cite{Ref:J.C:ASC, Ref:Prctc ,Ref:M.A:FOT, Ref:rethink,
Ref:Bid_MIT,Ref:exp} exploit the knowledge of self interference in
cancelation in the analog or digital domain. The residual interference still exists and can be modeled as Rayleigh fading~\cite{Ref:exp,Ref:Suppression,Ref:Inb_fd} when the direct link is effectively suppressed.

These suppression techniques can significantly reduce the
self-interference, which has made FD radios practically feasible in
the near future. This significant progress in FD has recently
inspired some very interesting work on FD signal processing.
In~\cite{Ref:Bid_MIT, Ref:Effs_est_err,
Ref:op_pc,Ref:Bi_use,Ref:ICC_pc, Ref:TRAPS}, the theoretical limits
of point to point bidirectional FD have been investigated by taking
into account the residual self-interference after suppression.
In~\cite{Ref:Bid_MIT}, the authors derived lower and upper bounds of
the achievable sum rate of bidirectional FD communications, and
proposed a transmission scheme to maximize the lower bound.
In~\cite{Ref:Bi_use}, the achievable sum rate of bidirectional FD
MIMO systems was analyzed and compared to the conventional HD MIMO systems over a spatial correlated channel. The
ergodic capacity of bidirectional FD transmission using one transmit
antenna and multiple receive antennas in the presence of channel
estimation error has been derived in~\cite{Ref:Effs_est_err}. a FD antenna mode selection scheme was investigated
in~\cite{Ref:TRAPS} for a simple $2\times2$ MIMO system, where each antenna is either configured as the transmit or receive antenna mode.
In~\cite{Ref:op_pc,Ref:ICC_pc}, the suboptimal and optimal dynamic power allocation
schemes were developed based on the sum rate maximization criterion.


In this paper, we consider a general FD MIMO system with $N_A$ and $N_B$ antennas equipped at two nodes. Such an FD MIMO system will create $N_AN_B$ possible links between the two FD MIMO nodes, with one possible link representing the channel from a transmit antenna of a node to a receive antenna of the other node. Since FD radios enable simultaneous bidirectional information exchanges between two FD MIMO nodes, a fundamental question arisen in such a system is how to select the link for each direction to optimize the system performance. In this paper we consider two performance metrics, weighted sum rate maximization and weighted sum symbol error rate (SER) minimization. The optimal\footnote{``Optimal'' in this paper means that this scheme can achieve the optimal performance under practical constraint, that only the distribution of the residual interference rather than the instantaneous one can be obtained at each node.} approach requires the exhaustive search from all possible
antenna links, however, as the number of antennas increases, such a brute-force search bears very high complexity in selection process.

To resolve this issue, in this paper we propose a simple Serial-Max selection algorithm by selecting the link with optimal performance for each direction in a two-step serial way, which can achieve asymptotically optimal performance. By using the law of total probability and order statics, the probability distribution functions of the two selected links are calculated, based on which, the closed-form expressions on average weighted sum rate and sum SER are derived. We show that the Serial-Max method approaches the
brute-force search method in terms of the average weighted sum rate and sum SER as the number of antennas
increases. The theoretical results are
verified by Monte-Carlo simulations.

The rest of this paper is organized as follows: Section II
introduces the system model. The proposed Serial-Max selection algorithm is presented in Section~III. Section~IV analyzes the performance of the Serial-Max method, including the average weighted sum rate and sum SER.
Simulation results are provided in Section V. In Section~VI, we draw
the main conclusions.



\section{System Model}
\begin{figure}[h!]
\centering
\includegraphics[width=4.2in]{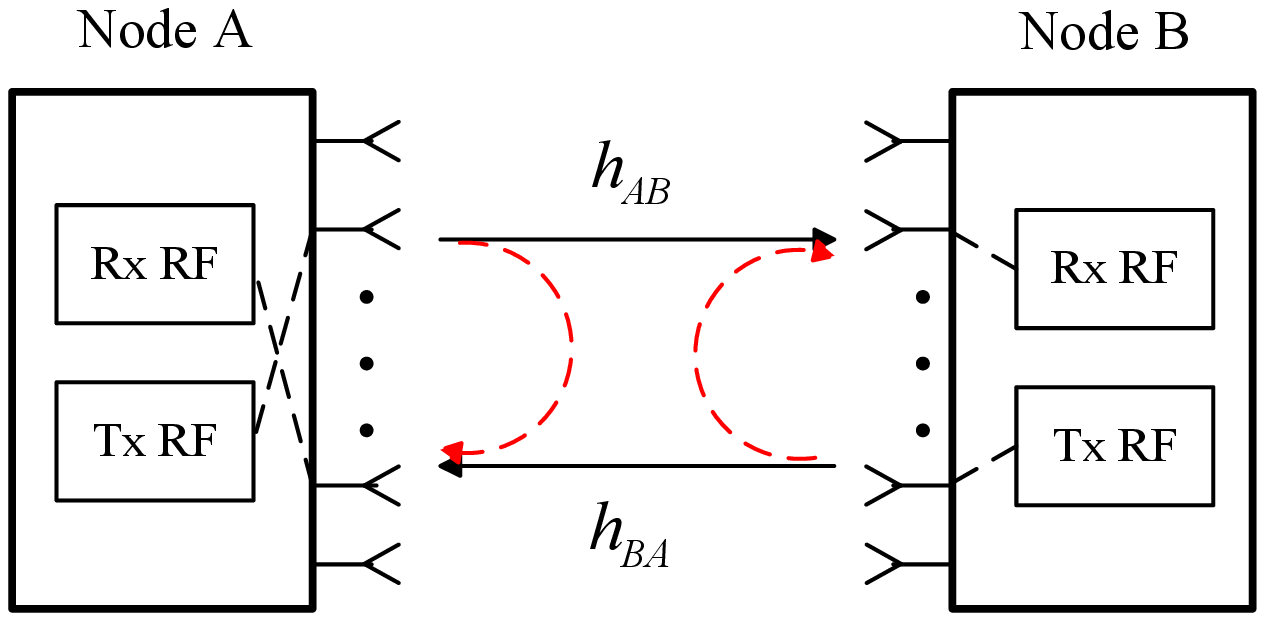}
\caption{Full duplex MIMO systems with simultaneous bidirectional link selection}
\label{fig:sm}
\end{figure}

In this paper, we consider a bidirectional communication scenario
between a pair of FD transceivers, node $A$ and $B$, as illustrated
in Fig.~\ref{fig:sm}, where node $A$ and $B$ are equipped with $N_A$ and $N_B$ antennas, respectively. Both nodes use the same
frequency band at the same time for FD operation. Each node employs
only one transmit and one receive RF chains, and any antenna can be configured to connect either the transmit or receive RF chain. In the
proposed simultaneous bidirectional link selection (SBLS) scheme, two antenna links are selected for simultaneous bidirectional communication by selecting a pair of transmit and receive antennas at both ends. Within each antenna pair, one antenna is selected for transmission and one is for reception.

We assume that the links between the two nodes are reciprocal and subject to
independent Rayleigh fading, and together denoted by a $N_A\times N_B$ channel
matrix ${\bf H}=\left[h^{ij}\right], i=1,2,...,N_A, j=1,2,...,N_B $. The
entry $h^{ij}$ represents the fading coefficient from the $i$-th antenna at node $A$ to
the $j$-th antenna at node $B$, and it follows the circularly symmetric Gaussian distribution
$\mathcal {C}\mathcal {N}(0,\sigma_h^2)$. All the possible communication channels are assumed to follow the non-selective independent
block fading, where the channel coefficients remain constant during a
time slot, and vary from one to another independently. In the beginning of each time slot, all the possible communication links can be estimated perfectly.

Since the two nodes are
operated in the FD mode, there exists self interference caused by
the signal leakage from the transmit antenna to the receive antenna
at the same node. We assume the passive propagation suppression and active analog/digital cancelation techniques are
employed to cancel the self interference. As indicated in~\cite{Ref:Suppression,Ref:Inb_fd}, the direct link of the self interference can be effectively suppressed, and the residual self interference can be approximated to follow the Rayleigh distribution~\cite{Ref:mul_outage,Ref:rly_slc}. In this paper, due to practical constraints, we assume that only the distribution of the residual interference, including the mean and variance, can be obtained and used in the selection process. Therefore, the selection process is based on the ``obtainable SINR''. This is equivalent to that based on the instantaneous SNR, which will be later shown that the former is a scaled version of the latter.


In the SBLS, the bidirectional links, i.e., $(I_T, J_R)$, from the $I_T$-th antenna at node $A$ to the $J_R$-th antenna at node $B$, and $(J_T, I_R)$, from the $J_T$-th antenna at node $B$ to the $I_R$-th antenna at node $A$, are selected. The
received signals at node $A$ and $B$, denoted by $y_A$ and $y_B$,
can be expressed as
\begin{align}\label{eq:sm_rcv_sgn}
{y_A} =& \sqrt {{P_t}} {h^{J_T I_R}}{x_B} + \sqrt { {P_t}} {h^{RI}_A}x_A +
{n_A}\\\nonumber {y_B} =& \sqrt {{P_t}} {h^{I_T J_R}}{x_A} + \sqrt {
{P_t}}h^{RI}_B {x_B} + {n_B},
\end{align}
where $h^{I_T J_R}$ and $h^{J_T I_R}$ denote the links corresponding to the
selected antenna pair from node $A$ to $B$ and that from node $B$ to $A$. $P_t$ is the
transmit power at each node. The second term denotes the residual
interference, at nodes $A$ and $B$, respectively. We assume
that both residual interference links, $h_A^{RI}$ and $h_B^{RI}$, are subject to Rayleigh fading
with zero mean and variance $\sigma_{RI}^2$. The AWGN at nodes
$A$ and $B$ are denoted by $n_A$ and $n_B$, which both follow
$\mathcal {C}\mathcal {N}(0,\sigma_n^2)$.

Without the knowledge of the instantaneous residual interference, the selection procedure is based on the ``obtainable SINR'' matrix ${\bf\Gamma}=[\gamma^{ij}]$. The entry in the ``obtainable SINR'' matrix is defined as $\gamma ^{ij}=\frac{\gamma^{ij}_s}{(\lambda_i+1)}$, where $\gamma^{ij}_s=\frac{P_t |h^{ij}|^2}{\sigma_n^2}$ is the instantaneous SNR, and $\lambda_i$ is the average INR. $\gamma^{ij}_s$ follows an exponential distribution with mean $\lambda_s=\frac{P_t\sigma_{h}^2}{\sigma_n^2}$, and the average INR is given by $\lambda_i=\eta\lambda_s=\frac{P_t\sigma_{RI}^2}{\sigma_n^2}$, where $\eta$ denotes the cancelation ability. The ``obtainable SINR'' $\gamma^{ij}$ is a scaled version of the instantaneous SNR $\gamma^{ij}_s$, because we assume that all the links have the same average INR $\lambda_i$.

In the following analysis in Section IV, we first calculate the instantaneous performance based on the instantaneous residual interference, and then average it with the distribution of fading channel and residual interference channel. We define the instantaneous SINR $\gamma_{AB}=\frac{\gamma_s^{I_T J_R}}{\gamma^{RI}_B+1}$ and $\gamma_{BA}=\frac{\gamma_s^{J_T I_R}} {\gamma^{RI}_A+1}$ for the two selected links. where the instantaneous INR $\gamma^{RI}_A$ and $\gamma^{RI}_B$ are both exponential random variables with mean $\lambda_i$.

\section{Simultaneous Bidirectional Link Selection (SBLS)}
In this section, we first introduce two optimal SBLS approaches to maximize the weighted sum rate and minimize the weighted sum SER, respectively, based on the ``obtainable SINR'' matrix, or equivalently the SNR matrix. Then, a low-complexity method which achieves asymptotically optimal performance, referred to as Serial-Max method, is proposed.


\subsection{SBLS based on Weighted Sum Rate Maximization Criterion (Max-WSR)}
In this subsection, we describe the SBLS based on Weighted Sum Rate Maximization criterion (Max-WSR) under the Gaussian input assumption. In this criterion, two communication links $\{ (I_T , J_R), (I_R, J_T ) \}$ from the $I_T$-th transmit antenna at node $A$ to the $J_R$-th receive antenna at
node $B$ and the $J_T$-th transmit antenna at node $B$ to the $I_R$-th
receive antenna at node $A$, are selected to maximize the weighted bidirectional sum rate
\begin{equation}\label{eq:AS_maxsr}
\left\{ {({I_T},{J_R}),({I_R},{J_T})} \right\} =
\mathop {\arg \max }\limits_{\scriptstyle {\kern 1.5pt}1 \le
{i_t},{i_r} \le N_A\hfill\atop{\scriptstyle 1\le {j_t},{j_r}\le
N_B\hfill\atop \scriptstyle {i_t} \ne {i_r},{j_t} \ne {j_r}\hfill}}
\left\{ wR\left({\gamma ^{{i_t}{j_r}}}\right)  + (1-w)R\left({{\gamma^{{i_r}{j_t}}}}\right)  \right\},
\end{equation}
where $R(\gamma)=\log_2\left[1+\gamma\right]$ denotes the rate under the ``obtainable SINR'' $\gamma=\frac{\gamma_s}{(\lambda_i+1)}$, and $0<w<1$ is the given weight of the transmission from node $A$ to $B$, depending on the rate requirement or quality of service (QoS) of each user. 


\subsection{SBLS Based on Weighted Sum SER Minimization (Min-WSER)}

In the SBLS based on weighted sum SER minimization criterion (Min-WSER) under the assumption that the input signal is modulated with finite constellations, the bidirectional antenna links are selected to minimize the weighted sum SER
\begin{equation}\label{eq:AS_minser}
\left\{ {({I_T},{J_R}),({I_R},{J_T})} \right\} =
\mathop {\arg \min }\limits_{\scriptstyle {\kern 1pt}1 \le
{i_t},{i_r} \le N_A\hfill\atop{\scriptstyle 1\le {j_t},{j_r}\le
N_B\hfill\atop \scriptstyle {i_t} \ne {i_r},{j_t} \ne {j_r}\hfill}}
\left\{
w{SE{R}\left( { { {{\gamma^{{i_r}{j_t}}}} }} \right)}  { + (1-w)SE{R}\left( { { {{\gamma^{{i_t}{j_r}}}}
}} \right)} \right\},
\end{equation}
where ${SE{R}\left( \gamma\right)}=\alpha Q\left( {\sqrt
{ \beta\gamma}}\right)$ represents the SER under the ``obtainable SINR'' $\gamma=\frac{\gamma_s}{(\lambda_i+1)}$.
$Q(\cdot )$ is the Gaussian $Q$-function~\cite{Ref:M.A:HOM}, and
$(\alpha,\beta)$ is a pair of constants determined by the modulation
format, e.g., $\alpha=1,\beta=2$ for BPSK modulation. 

\subsection{The Proposed Serial-Max SBLS}
The aforementioned SBLS schemes for the Max-WSR or Min-WSER criteria both require the brute-force search in order to find the optimal antenna pairs. This will become highly complex in selection process as the number of antennas increases. In this subsection, we introduce a low-complexity selection method, referred to as Serial-Max method, which selects the antenna pairs with maximum ``obtainable SINR'', or equivalently, the maximum SNR, in a two-step serial way.


In the first step, the best link with the maximum ``obtainable SINR'' is selected
\begin{equation}\label{eq:AS_grd1}
({I_1},{J_1}) = \mathop {\arg \max }\limits_{1 \le {i_1} \le
N_A,1\le{j_1}\le N_B} \left\{ {{\gamma ^{{i_1}{j_1}}}} \right\}.
\end{equation}
We use $\gamma^{1st} \buildrel\Delta \over = {\gamma^{{I_1}{J_1}}}$ to denote the ``obtainable SINR'' of the selected link $({I_1},{J_1})$, and $\gamma^{1st}_s$ to denote the SNR of this link. By removing the $I_1$-th column and $J_1$-th row from the ``obtainable SINR'' matrix $\bf \Gamma$ (and the corresponding SNR matrix $\bf \Gamma_s$),
we can obtain a $(N_A-1)\times(N_B-1)$ submatrix $\bf\Gamma'$ (and a corresponding pruned SNR matrix $\bf\Gamma'_s$).

In the second step, the link $(I_2,J_2)$ with the
maximum SNR is then selected from the pruned submatrix $\bf\Gamma'$ 
\begin{equation}\label{eq:AS_grd2}
({I_2},{J_2}) = \mathop {\arg \max }\limits_{\scriptstyle
{1 \le {i_2} \le
N_A,1\le{j_2}\le N_B}\hfill\atop \scriptstyle \hspace{5mm}{i_2} \ne
{I_1}, {j_2} \ne {J_1}\hfill} \left\{ {{\gamma ^{{i_r}{j_t}}}}
\right\}.
\end{equation}
We use $\gamma^{2nd} \buildrel \Delta \over = {\gamma^{{I_2}{J_2}}}$ to denote the ``obtainable SINR'' of the second selected link by the Serial-Max method, which is the maximum element of $\bf\Gamma'$, and we use $\gamma^{2nd}_s $ to denote the SNR of this link, which is the maximum element of the corresponding pruned SNR matrix $\bf\Gamma'_s$.

To maximize the weighted sum rate or minimize the weighted sum SER, the time-shared scheme can be employed, which allocates a fraction $\alpha$ of the time to use the best link for node $A$'s transmission, and the rest $1-\alpha$ to use the best link for node $B$'s transmission. Given $w$, the weighted maximization sum rate problem can be solved by calculating the allocation fraction $\alpha$. We have
\begin{equation}\label{eq:w_grd}
R = w\left[ {\alpha R\left( {\gamma ^{1st}} \right) + \left( {1 - \alpha } \right)R\left( {\gamma ^{2nd}} \right)} \right] + (1 - w)\left[ {\alpha R\left( {\gamma^{2nd}} \right) + \left( {1 - \alpha } \right)R\left( {\gamma ^{1st}} \right)} \right].
\end{equation}
To maximize $R$, we first calculate the derivative
\begin{equation}\label{eq:max_sr}
\frac{{\partial R}}{{\partial \alpha }} = \left(2w - 1\right)\left[R\left( {\gamma ^{1st}} \right) - R\left( {\gamma ^{2nd}} \right)\right].
\end{equation}
It is obvious that $R\left( {\gamma ^{1st}} \right) - R\left( {\gamma ^{2nd}} \right)$ is positive, therefore the allocation factor $\alpha$ depends on wether $w>0.5$ or not. Specifically, we have
\begin{equation}\label{eq:opt_alpha}
\left\{ \begin{array}{l}
\alpha  = 1,{\kern 27pt}  w > 0.5\\
\alpha  = 0,{\kern 27pt}  w < 0.5\\
0 \le \alpha  \le 1,{\kern 6pt} w = 0.5
\end{array} \right..
\end{equation}
Then, the weighted sum rate can be rewritten as
\begin{equation}\label{eq:wei_sum rate}
R=\max(w,1-w)R\left( {\gamma ^{1st}} \right)+\min(w,1-w){{ R}\left( {\gamma^{2nd}} \right)},
\end{equation}
The weighted sum SER minimization problem can be solved in a similar way, and we have
\begin{equation}\label{eq:wei_sum ser}
SER=\max(w,1-w)SER\left( {\gamma ^{1st}} \right)+\min(w,1-w){{ SER}\left( {\gamma ^{2nd}} \right)}.
\end{equation}
It is shown from (\ref{eq:wei_sum rate}) and (\ref{eq:wei_sum ser}) that the best link selected in the Serial-Max method will be allocated with the greater weight in the weighted sum rate and SER expression.

Regarding the performance of the Serial-Max method, we have the following Lemma.

\textbf{\emph{Lemma 1:}} The Serial-Max method can achieve the optimal weighted sum rate
and sum SER performances simultaneously, if the ``obtainable SINR'' of the second selected link $\gamma^{2nd}$ is the second or third largest element of the ``obtainable SINR'' matrix $\bf \Gamma$.

\emph{\textbf{Proof: }}We use $R_i$ and $SER_i$ to denote the respective rate and SER of the link with the $i$-th largest ``obtainable SINR''. If the second selected link corresponds to the second largest element in the ``obtainable SINR'' matrix $\bf\Gamma$, the weighted sum rate is given by $R=\max(w,1-w)R_1+\min(w,1-w)R_2$. It is obvious that in this case the Serial-Max method can achieve the optimal performance in terms of weighted sum rate. Similarly, it can be easily proved that the Serial-Max method can achieve the optimal performance of SER.


Recall that the pruned matrix $\bf\Gamma'$ is obtained by removing the row and column where the largest element, i.e., the first selected link, is located. Meanwhile, the second selected link is the largest element of the pruned matrix $\bf\Gamma'$. Therefore, if the second selected link corresponds to the third largest element in the original ``obtainable SINR'' matrix, it implies that the second largest element in $\bf\Gamma$ is removed in the aforementioned manipulation. In other words, the two links associated with the first and second largest elements in $\bf\Gamma$ share the same antenna, which can not be selected simultaneously. In this case, the largest and third largest elements is the best option for the two selected links. Therefore, the Serial-Max method which selects the largest and third largest elements in $\bf\Gamma$ achieves the optimal performances of weighted sum rate and sum SER. $\hfill \blacksquare$



Then, we have the following proposition.

\textbf{\emph{Proposition 1:}} The probability that the Serial-Max method does not
achieve the same performance as the optimal methods, denoted by $P_{not}$, is upper bounded by
\begin{equation}\label{eq:p_same}
P_{not}\le\frac{{\left( {{N_A} + {N_B} - 2} \right)\left( {{N_A} + {N_B} - 3} \right)}}{{\left( {{N_A}{N_B} - 1} \right)\left( {{N_A}{N_B} - 2} \right)}}.
\end{equation}

\emph{\textbf{Proof: }}According to Lemma 1, $P_{not}$ is upper bounded by the probability, denoted by $P_{not}^{2,3}$, that the ``obtainable SINR'' of the second selected link $\gamma^{2nd}$ is not the second nor third largest elements of the ``obtainable SINR'' matrix $\bf \Gamma$. This implies that both the second and third largest elements are in the same row or column as the largest element, and they are both removed in the process of obtaining the pruned ``obtainable SINR'' matrix $\bf\Gamma'$. Note that the elements of $\bf \Gamma$ are independent and identically distributed. Due to symmetry, each element of $\bf \Gamma$ has the same probability to be the $i$-th largest element. Thus, we have
\begin{equation}\label{eq:p_not23}
P_{not}^{2,3}=\frac{{\left( {{N_A} + {N_B} - 2} \right)\left( {{N_A} + {N_B} - 3} \right)}}{{\left( {{N_A}{N_B} - 1} \right)\left( {{N_A}{N_B} - 2} \right)}}.
\end{equation}
Combining the fact that $P_{not}$ is no more than $P_{not}^{2,3}$, Proposition 1 can be proved. $\hfill \blacksquare$



It is shown from (\ref{eq:p_same}) that the upper bound of $P_{not}$
decreases quadratically as the number of antennas increases, which implies the probability that the Serial-Max method selects the same pairs as the optimal one will increase, and thus approaches the optimal performance in terms of weighted sum rate and sum SER asymptotically.

In addition to the asymptotically-optimal performance, the complexity of the
Serial-Max method is much simpler than the exhaustive search
approach, as shown in Table~\ref{tab:1}. For the Serial-Max method,
in the first step, the maximum ``obtainable SINR'' is selected from a
$N_A\times N_B$ matrix, and
${N_{ A}\hspace{-0.3mm}N_{ B}}$
comparisons are required. Similarly, for the second step
$(N_{ A} - 1)(N_{ B} - 1)$
comparisons are needed, and the Serial method overall needs
$2{N_{ A}\hspace{-0.3mm}N_{ B}}-N_{ A} - N_{ B}+1$
comparisons. By contrary, the optimal method requires exhaustive
search in order to find the optimal antenna pairs, leading to
$\frac{N_AN_B(N_A-1)(N_B-1)}{2}$
comparisons. Therefore, the proposed Serial-Max algorithm can
approach the optimal algorithm with significantly reduced complexity.

\begin{table}[!hbp]
\centering \caption{Complexity comparison}\label{tab:1}
\setlength{\extrarowheight}{2mm}
\begin{tabular}{m{2.3cm}<{\centering}Im{5cm}<{\centering}Im{4.5cm}<{\centering}}    
\shline      
 &\vspace{-1.5mm}Optimal selection approach &\vspace{-1.5mm} Serial-Max approach\\
\shline
 \vspace{-1.5mm}{Complexity}\vspace{1mm} & \vspace{-1.2mm}{$\frac{N_AN_B(N_A-1)(N_B-1)}{2}$} & \vspace{-1.5mm}{$2{N_{ A}\hspace{-0.3mm}N_{ B}}-N_{ A} - N_{ B}+1$} \\       
\shline
\end{tabular}\\
\end{table}

\section{Performance Analysis}
The optimal selection methods are both very difficult to analyze,
and thus, in this section we analyze the performance of the
Serial-Max algorithm. It will be shown later in simulations that
the Serial-Max method can achieve near-optimal performance in
terms of average weighted sum rate and SER.

\subsection{Probability Distributions of Two Selected Links}
To analyze the performances of the Serial-Max method, the
distributions of the real instantaneous SINR corresponding to the two selected links $\gamma_{AB}$
and $\gamma_{BA}$ are required. For simplicity and without loss of generality, we consider the case of $w>0.5$ for the following analysis, which can be easily extended to $w\le 0.5$.

According to the description of the Serial-Max method
in~(\ref{eq:AS_grd1}), the SNR of the first selected link, $\gamma^{1st}_s$ is the largest
order statistic among ${N_{ A}\hspace{-0.3mm}N_{ B}}$ i.i.d. exponentially distributed random
variables $\gamma^{ij}_s$. The corresponding link $(I_1,J_1)$ is used for transmission from node $A$ to $B$. We have the following Lemma.

\textbf{\emph{Lemma 2:}} The CDF of $\gamma_{AB}$ can be given by
\begin{equation}\label{eq:prf_cdf_hab}
{F_{\gamma _{AB}^{}}}(x) = \sum\limits_{k = 0}^{{N_A}{N_B}} {\binom{N_AN_B}{k}{{\left( { - 1} \right)}^k}\frac{{e^{ - \frac{k}{{{\lambda _s}}}x}}}{{k\eta x + 1}}} .
\end{equation}

\emph{\textbf{Proof: }} The derivation is given in Appendix A.
$\hfill \blacksquare$

The second selected link in the Serial-Max method $(I_2,J_2)$ is used for transmission from node $B$ to $A$. According to (\ref{eq:AS_grd2}), we can obtain

\textbf{\emph{Lemma 3:}} The CDF expression of the instantaneous
link SINR of the second selected link, i.e., $\gamma_{BA}$, is given by
\begin{equation}\label{eq:cdf_hba}
{F_{{\gamma _{BA}}}}(x) = \sum\limits_{k = 1}^{{N_A} + {N_B} - 1} {\sum\limits_{l = {N_A}{N_B} - k}^{{N_A}{N_B}} {\sum\limits_{m = 0}^l {{{\left( { - 1} \right)}^m}{\mu _{k,l,m}}} \frac{{e^{ - \frac{{{N_A}{N_B} - l + m}}{{{\lambda _s}}}x}}}{{\left( {{N_A}{N_B} - l + m} \right)\eta x + 1}}} },
\end{equation}
where $\mu_{k,l,m}$ is expressed as
\begin{equation}
\mu_{ k,l,m} =
\frac{\binom{{N_{ A}\hspace{-0.3mm}N_{ B}}-k-1}{N_{ A} + N_{ B}-k-1}\binom{{N_{ A}\hspace{-0.3mm}N_{ B}}}{l}\binom{l}{m}}{\binom{{N_{ A}\hspace{-0.3mm}N_{ B}}-1}{N_{ A} + N_{ B}-2}}
.
\end{equation}

\textbf{\emph{Proof:}} The derivation is given in Appendix B.
$\hfill \blacksquare$

\subsection{Average weighted Sum Rate}

Based on the CDF expressions of the two selected links $\gamma_{AB}$ and $\gamma_{BA}$, in this section, the average weighted sum rate of the two links is obtained. Firstly, the instantaneous rate is calculated for given realizations of the selected communication channel and the corresponding self-interference channel. Then the average rate is obtained by averaging the instantaneous rate with respect to the distributions of the channels. Under the Gaussian input assumption, the average rate of the link with SINR $\gamma$ can be obtained as
\begin{align}\label{eq:rate}
\bar R&=\mathbb{E}\left[\log_2(1+ \gamma) \right]
\\\nonumber &=\frac{1}{{\ln 2}}\int\limits_0^\infty  {\frac{{1 -
{F_{\gamma }}(x)}}{{1 + x}}} dx ,
\end{align}
where $F_{\gamma}(x)$ is the CDF of $\gamma$.

\textbf{\emph{Proposition 2:}} The average weighted sum rate of the
Serial-Max method, denoted by $\bar R^{S-Max}$ is given by
\begin{equation}\label{eq:sum rate}
\bar R^{S-Max}=w\bar R_{AB}+(1-w){{\bar R}_{BA}},
\end{equation}
where $\bar R_{AB}$ and $\bar R_{BA}$ are the average rates of the
two selected links, respectively, which can be expressed
as
\begin{equation}\label{eq:bar_rate_ab}
{{\bar R}_{AB}} = \frac{1}{{\ln 2}}\sum\limits_{k = 0}^{{N_A}{N_B}} {\frac{{{{\left( { - 1} \right)}^k}\binom{N_AN_B}{k}}}{{1 - k\eta }}\left[ {{e^{\frac{1}{{\eta {\lambda _s}}}}}{{\rm{E}}_1}\left( {\frac{1}{{\eta {\lambda _s}}}} \right) - {e^{\frac{k}{{{\lambda _s}}}}}{{\rm{E}}_1}\left( {\frac{k}{{{\lambda _s}}}} \right)} \right]} ,
\end{equation}
and
\begin{equation}\label{eq:bar_rate_ba}
{{\bar R}_{B A}} = \frac{1}{{\ln 2}}\sum\limits_{k =
1}^{{N_A}+N_{B}
- 1} {\left( {{F_1} + {F_2}} \right)},
\end{equation}
where
\begin{equation}\label{eq:F1}
{F_1} = \sum\limits_{l = {N_A}{N_B} - k}^{{N_A}{N_B} - 1} {\sum\limits_{m = 0}^l {\frac{{{{\left( { - 1} \right)}^m}{\mu _{k,l,m}}}}{{1 - \left( {{N_A}{N_B} - l + m} \right)\eta }}\left[ {{e^{\frac{1}{{\eta {\lambda _s}}}}}{{\rm{E}}_1}\left( {\frac{1}{{\eta {\lambda _s}}}} \right) - {e^{\frac{{{N_A}{N_B} - l + m}}{{{\lambda _s}}}}}{{\rm{E}}_1}\left( {\frac{{{N_A}{N_B} - l + m}}{{{\lambda _s}}}} \right)} \right]} }  ,
\end{equation}
and
\begin{equation}\label{eq:F2}
{F_2} = \sum\limits_{m = 1}^{{N_A}{N_B}} {\frac{{{{\left( { - 1} \right)}^{m + 1}}{\mu _{k,{N_A}{N_B},m}}}}{{1 - m\eta }}\left[ {{e^{\frac{1}{{\eta {\lambda _s}}}}}{{\rm{E}}_1}\left( {\frac{1}{{\eta {\lambda _s}}}} \right) - {e^{\frac{m}{{{\lambda _s}}}}}{{\rm{E}}_1}\left( {\frac{m}{{{\lambda _s}}}} \right)} \right]}  .
\end{equation}
In addition, ${\rm E}_1(\cdot)$ denotes the exponential
integral function~\cite{Ref:M.A:HOM}.

\emph{\textbf{Proof:}} The derivation is given in Appendix C.
$\hfill \blacksquare$

It is shown from (\ref{eq:sum rate})--(\ref{eq:F2}) that the average weighted
sum rate of the Serial-Max method $\bar R^{S-Max}$ is only
determined by the average SNR $\lambda_s$ and cancelation ability $\eta$. When $\lambda_s$ goes to infinity, we have ${\rm E}_1(\varepsilon)\approx -\gamma-\ln(\varepsilon)$ and $e^\varepsilon \approx 1+\varepsilon$, for $\varepsilon=1/\lambda_s$, where $\gamma\approx 0.5772$ is Euler's constant~\cite{Ref:table}. After some manipulations, $\bar R^{S-Max}$ converges to a rate ceiling ${{\bar
R}^{S-Max}}_{\lambda\rightarrow\infty }$
\begin{align}\label{eq:sum_rate_infty}
\bar R_{\lambda_s\rightarrow\infty} ^{S - Max} \rightarrow \frac{w}{{\ln 2}}\sum\limits_{k = 0}^{{N_A}{N_B}} {\frac{{{{\left( { - 1} \right)}^k}\binom{N_AN_B}{k}\ln \left( {k\eta } \right)}}{{1 - k\eta }} + \frac{{1 - w}}{{\ln 2}}\sum\limits_{k = 1}^{{N_A} + {N_B} - 1} {\sum\limits_{m = 1}^{{N_A}{N_B}} {\frac{{{{\left( { - 1} \right)}^m}{\mu _{k,{N_A}{N_B},m}}\ln \left( {m\eta } \right)}}{{1 - m\eta }}} } } \\\nonumber
 + \frac{{1 - w}}{{\ln 2}}\sum\limits_{k = 1}^{{N_A} + {N_B} - 1} {\sum\limits_{l = {N_A}{N_B} - k}^{{N_A}{N_B} - 1} {\sum\limits_{m = 0}^l {\frac{{{{\left( { - 1} \right)}^m}{\mu _{k,l,m}}\ln \left[ {\left( {{N_A}{N_B} - l + m} \right)\eta } \right]}}{{1 - \left( {{N_A}{N_B} - l + m} \right)\eta }}} } }
\end{align}

\subsection{Average weighted sum SER}
In this section, we analyze the average weighted sum SER of the Serial-Max
method. For the SER analysis, we assume that the input signal is modulated with a finite constellation. Though the finite constellations like BPSK are used, the current self-interference cancellation technique can reduce the self-interference near to the noise level, as shown in~\cite{Ref:D.B:FD,Ref:Suppression}. It has thus been commonly assumed in many existing papers that the residual self-interference after cancellation follow the Rayleigh distributions~\cite{Ref:Inb_fd,Ref:mul_outage}. We also adopt this assumption in this paper. Firstly, the SER is calculated for a given set of channel realizations, similar to the weighted sum rate analysis. Then, the SER is averaged over the communication and self-interference channels. The average SER of the link with SINR $\gamma$ can be
written as
\begin{align}\label{eq:SER}
\overline {SER}  &= \alpha \mathbb{E}  \left[Q\left(\sqrt {\beta
\gamma } \right)\right]\\\nonumber &= \frac{\alpha\sqrt{\beta}
}{{2\sqrt {2\pi } }}\int\limits_0^\infty \frac{F_\gamma
\left(x\right)e^{ - \frac{{\beta x }}{2}} }{\sqrt{x}} dx.
\end{align}
$Q(\cdot)$ is the Gaussian Q-Function~\cite{Ref:M.A:HOM}.

In addition, if the first-order expansion of the PDF of $\gamma$ is
expressed as
\begin{equation}\label{eq:pdf_1st_order}
f_\gamma(x)=\frac{\zeta x^N}{\lambda^{N+1}}+o(x^{N+\varepsilon
}),
\end{equation}
the asymptotic SER can be obtained as~\cite{Ref:Li,Ref:SER}
\begin{equation}\label{eq:asypm_ser}
\overline {SER}  = \frac{{{2^N}\alpha \zeta \Gamma \left( {N +
\frac{3}{2}} \right)}}{{\sqrt \pi  \left( {N + 1} \right){{\left(
{\beta \lambda } \right)}^{N + 1}}}} + o\left( {\frac{1}{{{{\lambda }^{\left( {N + 1} \right)}}}}} \right).
\end{equation}

\emph{\textbf{Proposition 3:}} The average weighted sum SER of the Serial-Max
method, denoted by ${\overline {SER} ^{S-Max}}$ is
\begin{equation}\label{eq:ser_sum}
{\overline {SER} ^{S-Max}} = w{\overline {SER} _{AB}} + (1-w){\overline
{SER} _{BA}},
\end{equation}
where ${\overline {SER} _{AB}}$ and ${\overline {SER} _{BA}}$ are
the average SER of the two selected links, and can be calculated as
\begin{equation}\label{eq:ser_hab}
{\overline {SER} _{AB}} = \frac{1}{2} + \frac{{\alpha \sqrt {\beta \pi } }}{{\sqrt 2 }}\sum\limits_{k = 1}^{{N_A}{N_B}} {\frac{{{{( - 1)}^k}\binom{N_AN_B}{k}}}{{\sqrt {k\eta } }}{Q}\left( {\sqrt {\frac{2}{{\eta {\lambda _s}}} + \frac{\beta }{{k\eta }}} } \right)} {e^{\frac{1}{{\eta {\lambda _s}}} + \frac{\beta }{{2k\eta }}}} ,
\end{equation}
and
\begin{equation}\label{eq:ser_hba}
{\overline {SER} _{BA}} = \frac{{\alpha \sqrt {\beta \pi } }}{{\sqrt 2 }}\sum\limits_{k = 1}^{{N_A} + {N_B} - 1} {\left( {{G_1} + {G_2}} \right)} ,
\end{equation}
where
\begin{equation}\label{eq:G_1}
{G_1}\hspace{-0.7mm} =\hspace{-1mm} \sum\limits_{l = {N_A}{N_B} - k}^{{N_A}{N_B} - 1} {\sum\limits_{m = 0}^l {\frac{{{{\left( { - 1} \right)}^m}{\mu _{k,l,m}}}}{{\sqrt {\left( {{N_A}{N_B}\hspace{-0.7mm} -\hspace{-0.7mm} l \hspace{-0.7mm}+ m} \right)\eta } }}{Q}\hspace{-0.7mm}\left(\hspace{-0.7mm} {\sqrt {\frac{2}{{\eta {\lambda _s}}} \hspace{-0.7mm}+\hspace{-0.7mm} \frac{\beta }{{\left( {{N_A}{N_B} \hspace{-0.7mm}-\hspace{-0.7mm} l\hspace{-0.7mm} + m} \right)\eta }}} } \right)\hspace{-0.7mm}{e^{\frac{1}{{\eta {\lambda _s}}} + \frac{\beta }{{2\left( {{N_A}{N_B} - l + m} \right)\eta }}}}} } ,
\end{equation}
and
\begin{equation}\label{eq:G_2}
{G_2} = \sum\limits_{m = 1}^{{N_A}{N_B}} {\frac{{{{\left( { - 1} \right)}^m}{\mu _{k,{N_A}{N_B},m}}}}{{\sqrt {m\eta } }}{Q}\left( {\sqrt {\frac{2}{{\eta {\lambda _s}}} + \frac{\beta }{{m\eta }}} } \right){e^{\frac{1}{{\eta {\lambda _s}}} + \frac{\beta }{{2m\eta }}}}} .
\end{equation}

\emph{\textbf{Proof:}} The derivation is given in Appendix D. $\hfill
\blacksquare$

The SER performance converges to an error floor, when the average SNR $\lambda_s$ increases to infinity
\begin{align}\label{eq:ser_infty}
\overline {SER} _{\lambda\rightarrow\infty} ^{S - Max} \rightarrow& 1 - \frac{{w\alpha \sqrt {\beta \pi } }}{{\sqrt 2 }}\sum\limits_{k = 1}^{{N_A}{N_B}} {\frac{{{{( - 1)}^{k+1}}\binom{N_AN_B}{k}}}{{\sqrt {k\eta } }}{Q}\left( {\sqrt {\frac{\beta }{{k\eta }}} } \right){e^{\frac{\beta }{{2k\eta }}}}}\\\nonumber  &- \frac{{\left( {1 - w} \right)\alpha \sqrt {\beta \pi } }}{{\sqrt 2 }}\sum\limits_{k = 1}^{{N_A} + {N_B} - 1} {\frac{{{{\left( { - 1} \right)}^{m+1}}{\mu _{k,{N_A}{N_B},m}}}}{{\sqrt {m\eta } }}{Q}\left( {\sqrt {\frac{\beta }{{m\eta }}} } \right){e^{\frac{\beta }{{2m\eta }}}}}
\\\nonumber   &- \frac{{\left( {1 - w} \right)\alpha \sqrt {\beta \pi } }}{{\sqrt 2 }}\sum\limits_{k = 1}^{{N_A} + {N_B} - 1} {\sum\limits_{l = {N_A}{N_B} - k}^{{N_A}{N_B}} {\sum\limits_{m = 0}^l {\frac{{{{\left( { - 1} \right)}^{m+1}}{\mu _{k,l,m}}}}{{\sqrt {\left( {{N_A}{N_B} - l + m} \right)\eta } }}} } } \\\nonumber &{\hspace{5mm}}\times {Q}\left( {\sqrt {\frac{\beta }{{\left( {{N_A}{N_B} - l + m} \right)\eta }}} } \right){e^{\frac{\beta }{{2\left( {{N_A}{N_B} - l + m} \right)\eta }}}}.
\end{align}


On the other hand, when $\eta=0$, i.e. the self interference is
perfectly canceled, we can further calculate the asymptotic SER of the Serial-Max method at high SNR. Firstly, the CDF expression of
$\gamma_{AB}$ can be rewritten as
\begin{equation}\label{eq:CDF_AB_rewr}
{F_{{\gamma _{AB}}}}(x) = {\left( {1 - {e^{ - \frac{x}{{{\lambda _s}}}}}} \right)^{{N_A}{N_B}}}
\end{equation}
Then, the first order expansion of its corresponding PDF is given by
\begin{equation}
{f_{\gamma _{AB}^{}}}(x) = {N_A}{N_B}\frac{{{x^{{N_A}{N_B} - 1}}}}{{\lambda _s^{{N_A}{N_B}}}} + o\left( {{x^{{N_A}{N_B} - 1 + \varepsilon }}} \right).
\end{equation}
Using (\ref{eq:asypm_ser}), the asymptotic SER of
$\gamma_{AB}$ can be obtained
\begin{equation}\label{eq:asypm_ser_ab}
{\overline {SER} _{AB}} = \frac{{{u_1}}}{{\lambda _s^{{N_A}{N_B}}}} + o\left( {\frac{1}{{\lambda _s^{{N_A}{N_B}}}}} \right),
\end{equation}
where
\begin{equation}
u_1=\frac{{{2^{{{N_{ A}\hspace{-0.3mm}N_{ B}}} - 1}}\alpha \Gamma \left( {{{N_{ A}\hspace{-0.3mm}N_{ B}}} + \frac{1}{2}}
\right)}}{{{\beta ^{{{N_{ A}\hspace{-0.3mm}N_{ B}}}}}\sqrt \pi }}.
\end{equation}

Similarly, when the self interference cancelation is perfect, the
CDF of $\gamma_{BA}$ is rewritten as
\begin{equation}\label{eq:CDF_BA_rewr}
{F_{{\gamma _{BA}}}}(x) = \sum\limits_{k = 1}^{{N_A} + {N_B} - 1} {\sum\limits_{l = {N_A}{N_B} - k}^{{N_A}{N_B}} {{{\mu '}_{k,l}}} } {\left( {1 - {e^{ - \frac{x}{{{\lambda _s}}}}}} \right)^l}{e^{ - \frac{{\left( {{N_A}{N_B} - l} \right)x}}{{{\lambda _s}}}}} ,
\end{equation}
where
\begin{equation}
{{\mu
'}_{k,l}}=\frac{\binom{{N_{ A}\hspace{-0.3mm}N_{ B}}-k-1}{N_{ A} + N_{ B}-k-1}\binom{{N_{ A}\hspace{-0.3mm}N_{ B}}}{l}}{\binom{{N_{ A}\hspace{-0.3mm}N_{ B}}-1}{N_{ A} + N_{ B}-2}}.
\end{equation}
Then, the first order expansion of its corresponding PDF is
calculated as
\begin{equation}
{f_{\gamma _{BA}}}(x) = {{\mu
'}_{N_{ A} + N_{ B}
-
1,(N_{ A} - 1)(N_{ B} - 1)}}{(N_{ A} - 1)(N_{ B} - 1)}\frac{{{x^{(N_{ A} - 1)(N_{ B} - 1)
-
1}}}}{{\lambda_s^{(N_{ A} - 1)(N_{ B} - 1)}}}+
o\left(
{{x^{(N_{ A} - 1)(N_{ B} - 1)
+ \varepsilon }}} \right)
\end{equation}
Combing (\ref{eq:asypm_ser}), the asymptotic SER of
$\gamma_{BA}$ is
\begin{equation}\label{eq:asypm_ser_ba}
{\overline {SER} _{BA}} =
\frac{u_2}{{\lambda_s^{(N_{ A} - 1)(N_{ B} - 1)}}}
+ o\left(
{\frac{1}{{\lambda_s^{(N_{ A} - 1)(N_{ B} - 1)}}}}
\right),
\end{equation}
where
\begin{equation}
{u_2} = \frac{{{2^{{N_A}{N_B} - {N_A} - {N_B}}}\alpha \Gamma \left( {({N_A} - 1)({N_B} - 1) + \frac{1}{2}} \right)\binom{N_AN_B}{(N_A-1)(N_B-1)}}}{{{\beta ^{({N_A} - 1)({N_B} - 1)}}\sqrt \pi  \binom{N_AN_B-1}{N_A+N_B-2}}}.
\end{equation}

Combining (\ref{eq:asypm_ser_ab}) and (\ref{eq:asypm_ser_ba}), the
asymptotic weighted sum SER with perfect interference cancelation is
expressed as
\begin{equation}\label{eq:asypm_ser_sum}
{\overline {SER} ^{S-Max}} =
\frac{{(1-w)}{u_2}}{{\lambda_s^{(N_{ A} - 1)(N_{ B} - 1)}}}
+ o\left(
{\frac{1}{{\lambda_s^{(N_{ A} - 1)(N_{ B} - 1)}}}}
\right).
\end{equation}
It is implied by (\ref{eq:asypm_ser_sum}) that given $0<w<1$ the diversity order
of the Serial-Max method is
$(N_{ A} - 1)(N_{ B} - 1)$
with $\eta = 0$. This is coincident with a simple deduction of the existing result~\cite{Ref:AS_DMT}: The diversity order is determined by the worse link, i.e., the second best link which is selected from $\bf\Gamma'$ consisting of $(N_A-1)(N_B-1)$ i.i.d. elements. Moreover, the transmission direction with the greater weight will achieve the full diversity order of $N_AN_B$, and the other direction will achieve the diversity order of $(N_A-1)(N_B-1)$ for $w\ne 0.5$. On the condition that $w=0.5$, $\alpha$ can be a arbitrary fraction. Then the best link can be arbitrarily allocated to each direction. Therefore, the diversity orders of the two directions are both $(N_{ A} - 1)(N_{ B} - 1)$, and the achievable diversity is obviously $(N_{ A} - 1)(N_{ B} - 1)$ for the weighted sum SER.


\section{Simulation Results}
In this section, we provide the simulation results for our proposed
SBLS methods to validate the previous analysis. For simplicity, we consider that $N_A=N_B=N$ in the following simulations.

\subsection{Average Weighted Sum Rate}
\begin{figure} \centering
\includegraphics[width=5in]{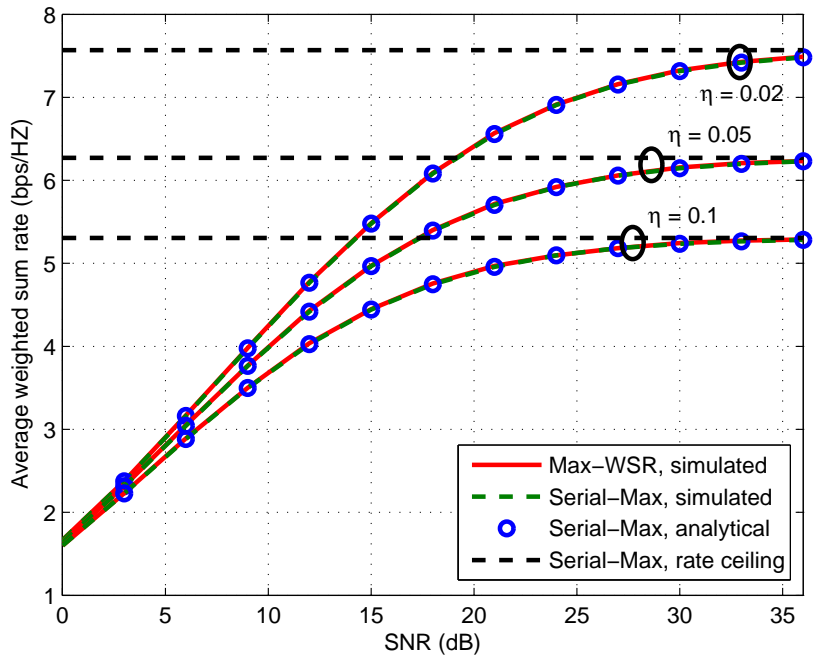}
\caption{Weighted sum rate performances of the Max-WSR and Serial-Max methods,
where $w=0.7$, $N=3$ and $\eta=0.02,0.05,0.1$.} \label{fig:Cap_N=3}
\end{figure}

Fig.~\ref{fig:Cap_N=3} depicts the average weighted sum rate of the Max-WSR
and Serial-Max methods versus SNR with different levels
of self interference $\eta=0.02,0.05,0.1$ for $N=3$ and $w=0.7$. It can be seen
that the weighted sum rate expression in (\ref{eq:sum rate}) perfectly
matches with the simulation results. In addition, the weighted sum rate
performance is limited by rate ceilings, which coincides with the
preceding analysis in (\ref{eq:sum_rate_infty}). From the figure, we
find that at low $\lambda_s$, the weighted sum rate performance for different
$\eta$ is quite similar, because the weighted sum rate performance at low
$\lambda_s$ is SNR-limited. However, at large $\lambda_s$, the residual self
interference will dominate the performance, and the performance is
limited by the rate ceiling caused by the residual self interference. The figure also reveals that the Serial-Max method
achieves almost the same average weighted sum rate as the Max-WSR method
across all SNR regions.

\begin{figure} \centering
\includegraphics[width=5in]{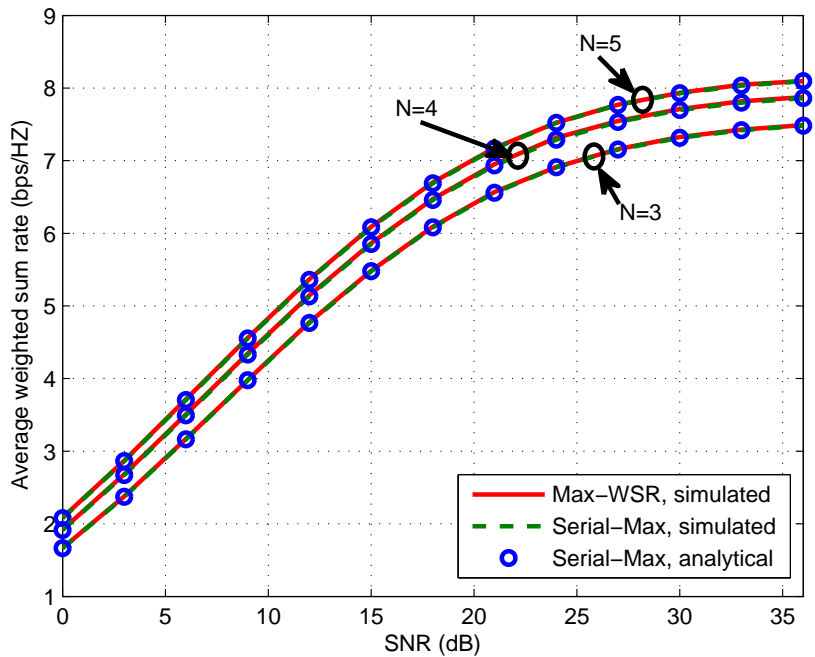}
\caption{Weighted sum rate performances of the Max-WSR and Serial-Max methods,
where $w=0.7$, $\eta=0.02$ and $N=3,4,5$.} \label{fig:Cap_eta=0.05}
\end{figure}

In Fig.~\ref{fig:Cap_eta=0.05}, we illustrate the average weighted sum rate
of the Max-WSR and Serial-Max methods for different numbers of
antennas $N=3,4,5$ where the self interference cancelation
coefficient is $\eta=0.02$. It can be observed that, for different
numbers of antennas, the Serial-Max algorithm achieves almost the
same average weighted sum rate as the Max-WSR one across all SNR region.
We can also find from this figure that the
average weighted sum rate increases with the number of antennas.

\subsection{Average Weighted Sum SER}
\begin{figure} \centering
\includegraphics[width=5in]{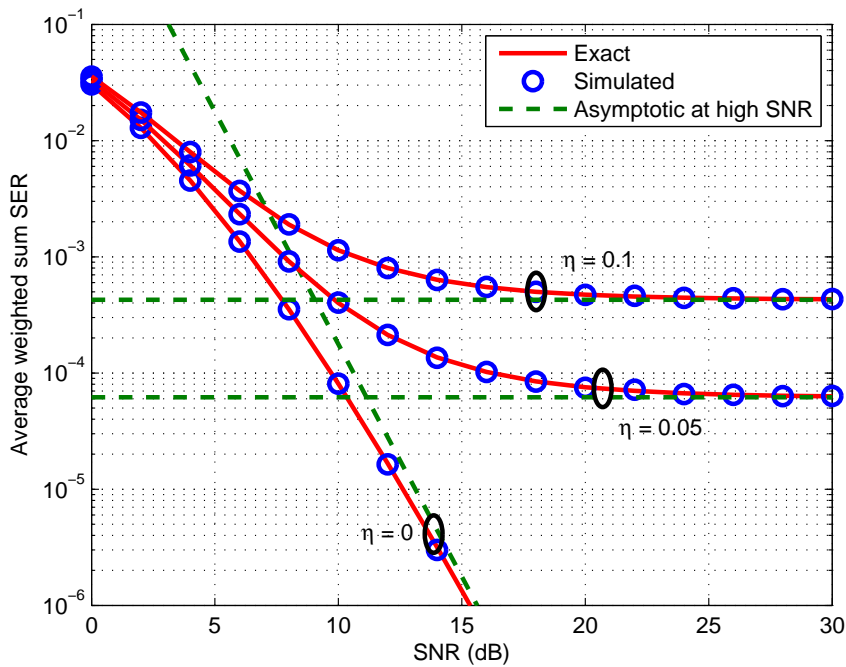}
\caption{Weighted sum SER performances of the Serial-Max method, where $N=3$
and $\eta=0,0.05,0.1,0.5$.} \label{fig:ser_N=3}
\end{figure}

The following simulations of weighted sum SER performance are conducted with BPSK modulation. In Fig.~\ref{fig:ser_N=3}, the weighted sum SER performance of the Serial-Max
method is provided for different $\eta=0,0.05, 0.1, 0.5$, where
large $\eta$ means severe self interference whereas small $\eta$
means slight self interference level. Especially, $\eta = 0$ means
perfect interference cancelation. This figure verifies the weighted sum SER
expression given by Proposition 3. Based on the figure, it can be
observed that the simulated SER performance for $\eta=0$ tightly
matches with the asymptotic one given by (\ref{eq:asypm_ser_sum}) at
high SNR, while the SER performances for $\eta> 0 $ are
constrained by error floors evaluated by (\ref{eq:ser_infty}) at
high SNR. It can also be seen that when the self interference is
perfectly canceled, the diversity order of the Serial-Max method is
$(N_{ A} - 1)(N_{ B} - 1)$.
However when the residual self interference exists, the Serial-Max
method has a zero diversity order.

\begin{figure} \centering
\includegraphics[width=5in]{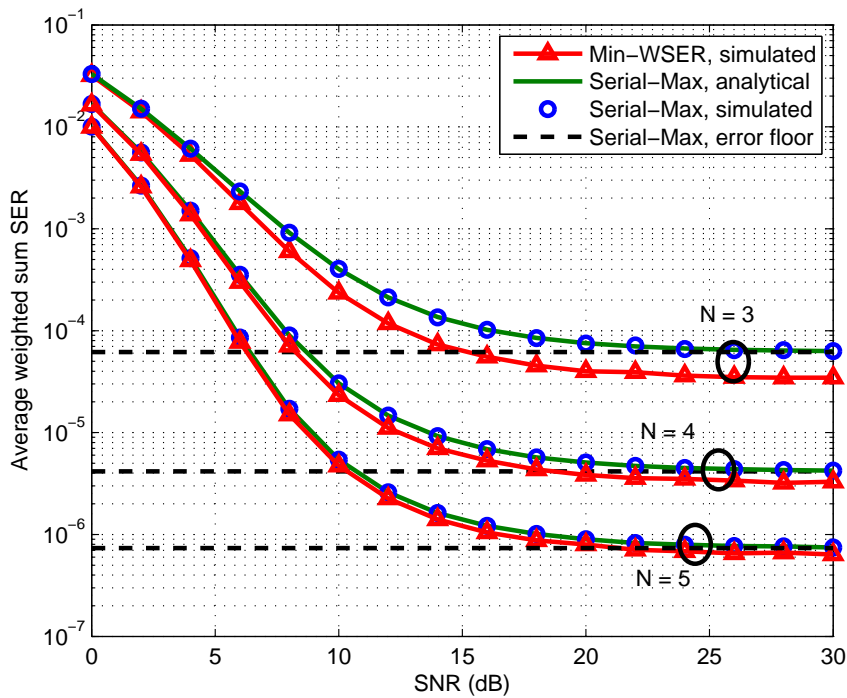}
\caption{Weighted sum SER performances of the Min-WSER and Serial-Max methods,
where $\eta=0.05$ and $N=3,4,5$.} \label{fig:ser_eta=0.1}
\end{figure}

Fig.~\ref{fig:ser_eta=0.1} shows the average weighted sum SER for different
numbers of antennas $N=3,4,5$ where the self interference level
$\eta=0.1$ is assumed. It shows that both the Min-WSER and Serial-Max methods are
limited by error floors at high SNR due to the residual self
interference. As the number of antennas increase, the SER
performances of both methods including the error floor are improved.
Moreover, the Serial-Max method performs closer to the Min-WSER scheme as $N$ increases.


\begin{figure} \centering
\includegraphics[width=5in]{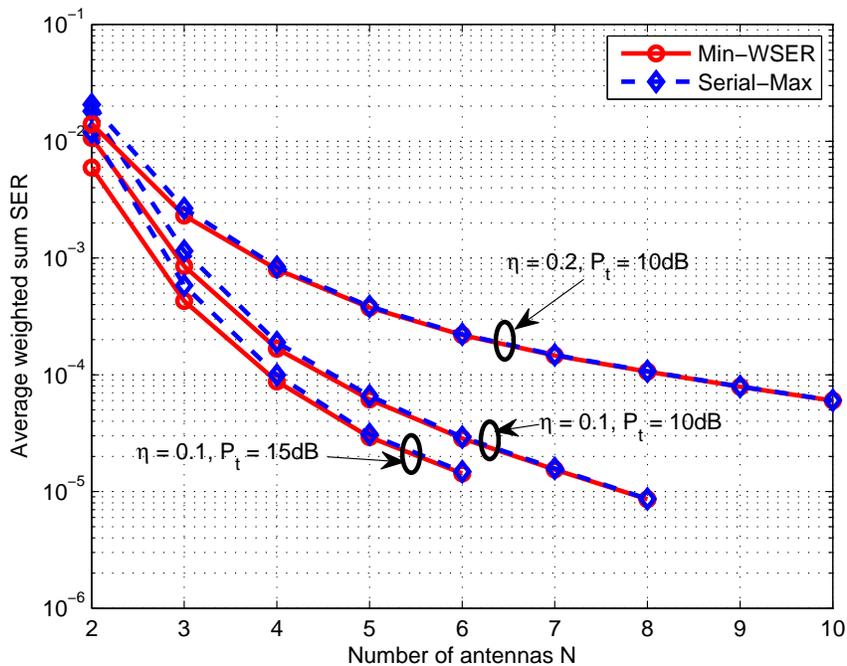}
\caption{Simulated weighted sum SER performances of the Min-WSER and Serial-Max
methods, where $\eta=0.1,0.2$ and $P_t=10, 15 \rm dB$.}
\label{fig:ser_N}
\end{figure}

Fig.~\ref{fig:ser_N} compares the simulated weighted sum SER performance of
the Min-WSER and Serial-Max methods with different number of
antennas. Combinations of different SNR and self
interference levels ($\lambda_s= 10, 15 \rm dB$, $\eta = 0.1, 0.2$) are
provided. It can be observed that the gaps between the Min-WSER and
Serial-Max methods are reduced as $N$ increases. It also shows that
the SER performance of the Serial-Max method approaches the
Min-WSER method when the self interference is large or SNR is small. This is because in these cases these two factors
dominate the SER performances of both methods.

\subsection{Computational complexity}
\begin{figure} \centering
\includegraphics[width=5in]{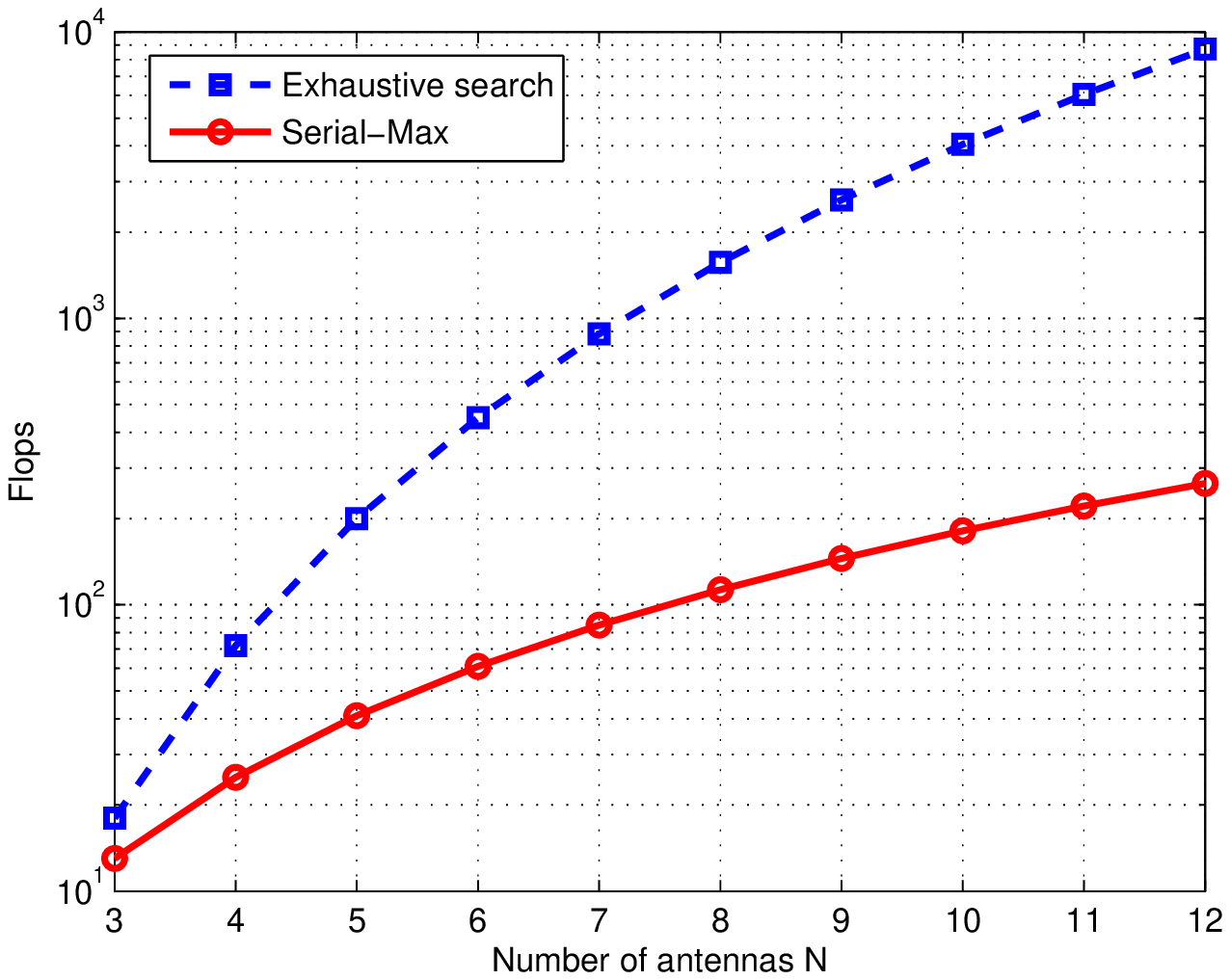}
\caption{Complexity comparison between the optimal method and
the Serial-Max method.} \label{fig:complexity}
\end{figure}

To compare the computational complexity of the optimal and
Serial-Max methods, the number of required floating-point operations
(flops) are presented in Fig.~\ref{fig:complexity}. It is clear that
the optimal method has a high complexity due to the brute-force search operation, while the Serial-Max method can
provide significant a complexity reduction, especially when the number
of antennas is large.

\section{Conclusions}
In this paper, we have proposed an opportunistic bidirectional link selection approach in FD MIMO systems.
The optimal approach based on the ``obtainable SINR'', defined as the ratio of the instantaneous SNR and average INR plus one, requires exhaustive search, bearing high complexity. To facilitate the selection process and reduce
the computational complexity, a simple Serial-Max method with near
optimal performances was proposed. The performance analysis was provide for the Serial-Max method, including the weighted sum rate and SER performances. It was shown that the
proposed Serial-Max method approaches the respective weighted sum rate and weighted sum
SER performances of the exhaustive search methods when the number of antennas increases.

\section*{Appendix A: Proof of Lemma~2}
According to the description of the Serial-Max method, the CDF of $\gamma_{AB}$ is expressed as
\begin{align}\label{eq:cdf_r_ab}
{F_{{\gamma _{AB}}}}(x) &= \Pr \left( {\frac{{\gamma _s^{{I_1}{J_1}}}}{{\gamma _B^{RI} + 1}} < x} \right)\\\nonumber
& = \Pr \left( {\gamma _s^{{I_1}{J_1}} < x\left( {\gamma _B^{RI} + 1} \right)} \right),
\end{align}
where $\gamma^{{I_1}{J_1}}_s$ is the largest one of $N_A \times N_B$ i.i.d. exponential-distributed random variables, and its CDF is given by~\cite{Ref:H.A:OSt}
\begin{equation}\label{eq:cdf_r_1st}
{F_{\gamma _s^{{I_1}{J_1}}}}(x) = {\left( {1 - {e^{ - \frac{x}{{{\lambda _s}}}}}} \right)^{{N_A}{N_B}}}.
\end{equation}
$\gamma^{RI}_B$ follows exponential distribution with average $\lambda_i$. Substituting the CDF expression of $\gamma^{{I_1}{J_1}}_s$ and $\gamma^{RI}_B$,~(\ref{eq:cdf_r_ab}) can be calculated as
\begin{equation}\label{eq:cdf_r_ab1}
{F_{{\gamma _{AB}}}}(x) = \int\limits_0^\infty  {{{\left[ {1 - {e^{ - \frac{{x\left( {\gamma _B^{RI} + 1} \right)}}{{{\lambda _s}}}}}} \right]}^{{N_A}{N_B}}}\frac{{{e^{ - \frac{{\gamma _B^{RI}}}{{{\lambda _i}}}}}}}{{{\lambda _i}}}d\gamma _B^{RI}}.
\end{equation}
Through some manipulations, we can obtain the CDF of $\gamma_{AB}$.

\section*{Appendix B: Proof of Lemma~3}
As aforementioned, we have $\gamma_{BA}={\gamma _s^{{I_2}{J_2}}}/({\gamma _A^{RI} + 1})$, and its CDF expression is
\begin{align}\label{eq:cdf_r_ba}
{F_{{\gamma _{BA}}}}(x) &= \Pr \left( {\frac{{\gamma _s^{{I_2}{J_2}}}}{{\gamma _A^{RI} + 1}} < x} \right)\\\nonumber
& = \Pr \left( {\gamma _s^{{I_2}{J_2}} < x\left( {\gamma _A^{RI} + 1} \right)} \right).
\end{align}
$\gamma _s^{{I_2}{J_2}}$ is the maximum order statistic in the pruned matrix ${\bf\Gamma}_s'$ consisting of $(N_A-1) \times (N_B-1)$
random variables following the exponential distribution. Since ${\bf\Gamma}_s'$ is obtained by removing
the maximum element and other
$N_{ A} + N_{ B}-2$
elements from ${\bf\Gamma}_s$, $\gamma _s^{{I_2}{J_2}}$ could possibly be
one of any
$({N_{ A}\hspace{-0.3mm}N_{ B}}-k)$-th
order statistic of ${\bf\Gamma}_s$,
$k=\{1,2...,N_{ A} + N_{ B}-1\}$.
As a result, the CDF of ${\gamma _s^{{I_2}{J_2}}}$ is given by
\begin{equation}\label{eq:prf_cdf_hba_1}
{F_{\gamma _s^{{I_2}{J_2}}}}(x) = \sum\limits_{k =
1}^{N_{ A} + N_{ B}
- 1}
{{p_k}{F_{\gamma}^{({{N_{ A}\hspace{-0.3mm}N_{ B}}}
- k)}}(x)},
\end{equation}
where $p_k$ is the probability that ${\gamma _s^{{I_2}{J_2}}}$ is the
$({N_{ A}\hspace{-0.3mm}N_{ B}}-k)$-th order statistic in ${\bf\Gamma}_s$, and
${F_{\gamma}^{({{N_{ A}\hspace{-0.3mm}N_{ B}}} - k)}}(x)$ represents the CDF of the
$({N_{ A}\hspace{-0.3mm}N_{ B}}-k)$-th order statistic among ${N_{ A}\hspace{-0.3mm}N_{ B}}$ variables.

For $n$ i.i.d. variables, each with CDF $F(x)$, the CDF of the
$r$-th largest order statistic can be written as
\begin{equation}\label{eq:cdf_r_os}
{F^{(r)}}(x) = \sum\limits_{i = r}^n {{\binom{n}{i}}{F^i}(x){{\left[
{1 - F(x)} \right]}^{n - i}}}.
\end{equation}
Here, they are all exponential variables with average $\lambda_s$. Then, the CDF of the
$({N_{ A}\hspace{-0.3mm}N_{ B}}-k)$-th order statistic can be expressed as
\begin{align}\label{eq:cdf_os_n2-k}
F_\gamma ^{({{N_{ A}\hspace{-0.3mm}N_{ B}}} - k)}(x) =& \hspace{-2mm}\sum\limits_{l = {{N_{ A}\hspace{-0.3mm}N_{ B}}} - k}^{{{N_{ A}\hspace{-0.3mm}N_{ B}}}}
{\hspace{-1mm}\binom{{N_{ A}\hspace{-0.3mm}N_{ B}}}{l} {{\left( {1 - {e^{ - \frac{x}{{\bar \gamma }}}}}
\right)}^l}\hspace{-0.3mm}{e^{ - \frac{{{{N_{ A}\hspace{-0.3mm}N_{ B}}} - l}}{{\bar \gamma }}x}}}\\\nonumber
=&\hspace{-2mm} \sum\limits_{l = {{N_{ A}\hspace{-0.3mm}N_{ B}}} - k}^{{{N_{ A}\hspace{-0.3mm}N_{ B}}}} \hspace{1mm} \sum\limits_{m = 0}^l
\binom{{N_{ A}\hspace{-0.3mm}N_{ B}}}{l}\binom{l}{m}{{( - 1)}^m}e^{ - \frac{{{{N_{ A}\hspace{-0.3mm}N_{ B}}} - l +
m}}{{\bar \gamma }}x} .
\end{align}

On the other hand, if ${\gamma _s^{{I_2}{J_2}}}$ is the
$({N_{ A}\hspace{-0.3mm}N_{ B}}-k)$-th
order statistic in ${\bf\Gamma}_s$, it means that among the
$(N_{ A} + N_{ B}-1)$
elements removed from the matrix ${\bf\Gamma}_s$, there are $k$ elements
larger than ${\gamma _s^{{I_2}{J_2}}}$. To satisfy that the largest
element of ${\bf\Gamma}_s'$ is the $(k+1)$ largest element in
${\bf\Gamma}_s$, we can first remove the first $k$ largest elements in
the matrix ${\bf\Gamma}_s$, and then, keeping the $(k+1)$-th largest
element unremoved, we remove the other
$(N_{ A} + N_{ B}-1-k)$
elements randomly. Thus, $p_k$ is calculated as
\begin{equation}\label{eq:p_k}
p_k=\frac{\binom{{N_{ A}\hspace{-0.3mm}N_{ B}}-k-1}{N_{ A} + N_{ B}-k-1}}{\binom{{N_{ A}\hspace{-0.3mm}N_{ B}}-1}{N_{ A} + N_{ B}-2}},
\end{equation}
where the numerator implies that when we remove
$(N_{ A} + N_{ B}-1-k)$
elements randomly from
$({N_{ A}\hspace{-0.3mm}N_{ B}}-k-1)$
elements, there are
$\binom{{N_{ A}\hspace{-0.3mm}N_{ B}}-k-1}{N_{ A} + N_{ B}-k-1}$
possibilities, while the denominator means that when we remove
$(N_{ A} + N_{ B}-2)$
elements (besides ${\gamma _s^{{I_1}{J_1}}}$, the largest one of
${\bf\Gamma}_s$) randomly from
$({N_{ A}\hspace{-0.3mm}N_{ B}}-1)$
elements, there are
$\binom{{N_{ A}\hspace{-0.3mm}N_{ B}}-1}{N_{ A} + N_{ B}-2}$
possibilities.

Finally, substituting (\ref{eq:cdf_os_n2-k}) and (\ref{eq:p_k}) into
(\ref{eq:prf_cdf_hba_1}), the CDF expression of
${\gamma _s^{{I_2}{J_2}}}$ can be obtained as
\begin{equation}\label{eq:cdf_r_2nd}
{F_{\gamma _s^{{I_2}{J_2}}}}(x) = \sum\limits_{k = 1}^{{N_A} + {N_B} - 1} {\sum\limits_{l = {N_A}{N_B} - k}^{{N_A}{N_B}} {\sum\limits_{m = 0}^l {{{\left( { - 1} \right)}^m}{\mu_{k,l,m}}} {e^{ - \frac{{{N_A}{N_B} - l + m}}{{\lambda }}x}}} } ,
\end{equation}
where $\mu_{ k,l,m}$ is expressed as
\begin{equation}
\mu_{ k,l,m} =
\frac{\binom{{N_{ A}\hspace{-0.3mm}N_{ B}}-k-1}{N_{ A} + N_{ B}-k-1}\binom{{N_{ A}\hspace{-0.3mm}N_{ B}}}{l}\binom{l}{m}}{\binom{{N_{ A}\hspace{-0.3mm}N_{ B}}-1}{N_{ A} + N_{ B}-2}}
.
\end{equation}
Then, substituting (\ref{eq:cdf_r_2nd}) into (\ref{eq:cdf_r_ba}) and combining the CDF of $\gamma_A^{RI}$ we have
\begin{equation}\label{eq:cdf_r_ba1}
{F_{{\gamma _{BA}}}}(x) = \int\limits_0^\infty  {\sum\limits_{k = 1}^{{N_A} + {N_B} - 1} {\sum\limits_{l = {N_A}{N_B} - k}^{{N_A}{N_B}} {\sum\limits_{m = 0}^l {{{\left( { - 1} \right)}^m}{u_{k,l,m}}} {e^{ - \frac{{{N_A}{N_B} - l + m}}{{{\lambda _i}}}x\left( {\gamma _A^{RI} + 1} \right)}}} } \frac{{{e^{ - \frac{{\gamma _A^{RI}}}{{{\lambda _i}}}}}}}{{{\lambda _i}}}d\gamma _A^{RI}}.
\end{equation}
Through some manipulations, Lemma 3 can be obtained.

\section*{Appendix C: Proof of Proposition~2}

Combining the CDF expression of $\gamma_{AB}$ in (\ref{eq:prf_cdf_hab}) and (\ref{eq:rate}),
we can obtain
\begin{align}\label{eq:r_ab2}
{{\bar R}_{AB}} &= \frac{1}{{\ln 2}}\sum\limits_{k = 1}^{{N_A}{N_B}} {{{\left( { - 1} \right)}^{k + 1}}\binom{N_AN_B}{k}\int\limits_0^\infty  {\frac{{{e^{ - \frac{k}{{{\lambda _s}}}x}}}}{{\left( {k\eta x + 1} \right)\left( {1 + x} \right)}}} dx} \\\nonumber &= \frac{1}{{\ln 2}}\sum\limits_{k = 1}^{{N_A}{N_B}} {{{\left( { - 1} \right)}^{k + 1}}\binom{N_AN_B}{k}\int\limits_0^\infty  {\frac{1}{{1 - k\eta }}\left[ {\frac{{{e^{ - \frac{k}{{{\lambda _s}}}x}}}}{{\left( {1 + x} \right)}} - \frac{{k\eta {e^{ - \frac{k}{{{\lambda _s}}}x}}}}{{k\eta x + 1}}} \right]dx} }  .
\end{align}
Using the integral ${{\rm{E}}_1}\left( x \right) = {e^{ - x}}\int_0^\infty  {\frac{{{e^{ - t}}}}{{t + x}}dt} $~\cite{Ref:M.A:HOM}, (\ref{eq:r_ab2}) can be given by
\begin{equation}\label{eq:r_ab3}
{{\bar R}_{AB}}= \frac{1}{{\ln 2}}\sum\limits_{k = 1}^{{N_A}{N_B}} {\frac{{{{\left( { - 1} \right)}^k}
\binom{N_AN_B}{k}}}{{1 - k\eta }}\left[ {{e^{\frac{1}{{\eta {\lambda _s}}}}}{{\rm{E}}_1}\left( {\frac{1}{{\eta {\lambda _s}}}} \right) - {e^{\frac{k}{{{\lambda _s}}}}}{{\rm{E}}_1}\left( {\frac{k}{{{\lambda _s}}}} \right)} \right]}
\end{equation}

On the other hand, the CDF of ${\gamma _{BA}}$ in
(\ref{eq:cdf_hba}) can be rewritten as
\begin{equation}\label{eq:cdf_hba_re}
{F_{\gamma _{BA}}}(x) = 1 - \sum\limits_{k =
1}^{N_{ A} + N_{ B}
- 1} {\left( {{f_1} + {f_2}} \right)},
\end{equation}
where
\begin{equation}\label{eq:f11}
{f_1} = \sum\limits_{l = {N_A}{N_B} - k}^{{N_A}{N_B} - 1} {\sum\limits_{m = 0}^l {{{\left( { - 1} \right)}^{m + 1}}{\mu _{k,l,m}}\frac{{e^{ - \frac{{{N_A}{N_B} - l + m}}{{{\lambda _s}}}x}}}{{\left( {{N_A}{N_B} - l + m} \right)\eta x + 1}}} } ,
\end{equation}
and
\begin{equation}\label{eq:f22}
{f_2} = \sum\limits_{m = 1}^{{N_A}{N_B}} {{{\left( { - 1} \right)}^{m + 1}}{\mu _{k,{N_A}{N_B},m}}\frac{{{e^{ - \frac{m}{{{\lambda _s}}}x}}}}{{m\eta x + 1}}}  .
\end{equation}
Then, substituting (\ref{eq:cdf_hba_re})--(\ref{eq:f22}) into
(\ref{eq:rate}), we have
\begin{equation}\label{eq:sr_ba}
{{\bar R}_{BA}} = \frac{1}{{\ln 2}}\sum\limits_{k =
1}^{N_{ A} + N_{ B}
- 1} {\left( {\int\limits_0^\infty  {\frac{{{f_1} + {f_2}}}{{1 +
x}}dx} } \right)}.
\end{equation}
After some manipulations, the average rate $\bar R_{BA}$ in
(\ref{eq:bar_rate_ba}) can be obtained. Finally, combining the
expression of $\bar R_{AB}$ and $\bar R_{BA}$, Proposition 2 can be
proved.

\section*{Appendix D: Proof of Proposition~3}

Firstly, substituting the CDF of
$\gamma_{AB}$ in (\ref{eq:prf_cdf_hab}) into (\ref{eq:SER}),
we have
\begin{equation}\label{eq:ser_hab_1}
{\overline {SER} _{AB}} = \frac{1}{2}-\frac{{\alpha \sqrt \beta  }}{{2\sqrt {2\pi } }}\sum\limits_{k = 1}^{{N_A}{N_B}} {{{( - 1)}^{k+1}}\binom{N_AN_B}{k}\int\limits_0^\infty  {\frac{{{e^{ - \left( {\frac{\beta }{2} + \frac{k}{{{\lambda _s}}}} \right)x}}}}{{\left( {k\eta x + 1} \right)\sqrt x }}dx} } .
\end{equation}
Applying Integration by substitution, the integral part is rewritten as
\begin{equation}\label{eq:int_subst}
\int\limits_0^\infty  {\frac{{{e^{ - \left( {\frac{\beta }{2} + \frac{k}{{{\lambda _s}}}} \right)x}}}}{{\left( {k\eta x + 1} \right)\sqrt x }}dx} \mathop  = \limits^{t = \sqrt x } \int\limits_0^\infty  {\frac{{2{e^{ - \left( {\frac{\beta }{2} + \frac{k}{{{\lambda _s}}}} \right){t^2}}}}}{{\left( {k\eta {t^2} + 1} \right)t}}dt}.
\end{equation}
Using the integral $\int_0^\infty  {\frac{{{e^{ - {a^2}{x^2}}}}}{{{x^2} + {b^2}}}dx}  = Q\left( {\sqrt 2 ab} \right)\frac{\pi }{b}{e^{{a^2}{b^2}}}$~\cite{Ref:table}, (\ref{eq:int_subst}) can be calculated as
\begin{equation}\label{eq:intgral}
 \int\limits_0^\infty  {\frac{{{e^{ - \left( {\frac{\beta }{2} + \frac{k}{{{\lambda _s}}}} \right)x}}}}{{\left( {k\eta x + 1} \right)\sqrt x }}dx}  = \frac{{2\pi }}{{\sqrt {k\eta } }}Q\left( {\sqrt {\frac{2}{{\eta {\lambda _s}}} + \frac{\beta }{{k\eta }}} } \right){e^{\frac{1}{{\eta {\lambda _s}}} + \frac{\beta }{{2k\eta }}}}.
\end{equation}
Substituting (\ref{eq:intgral}) into (\ref{eq:ser_hab_1}), and after some manipulation, (\ref{eq:ser_hab})
can be obtained.

Then, the average SER of $\gamma_{BA}$ in (\ref{eq:ser_hba}) can
be obtained by substituting (\ref{eq:cdf_hba}) into (\ref{eq:SER}) and through the similar manipulations. Finally, combining the two
average SER expressions, Proposition~3 can be proved.

\end{document}